\documentclass[prb,preprint,showpacs]{revtex4}
\usepackage{mathrsfs}
\usepackage{graphicx}
\usepackage{subeqnarray}
\usepackage{color}
\begin{document}
\title{{Tuning the quantum tunneling and quantum information properties among the trion states in coupled quantum dots}}
\author{Y.B. Wen, and W.X. Yan\footnote{To whom correspondence should be addressed,
E-mail: {yanwxsxu@yahoo.com.cn}}} \affiliation{Department of
Physics, College of Physics and Electronics, Shanxi University,
Taiyuan, 030006, People's republic of China}

\begin{abstract}
Quantum transition and information properties of the Coulomb coupled
trion (electron-electron-hole) in the double quantum dots under the
influence of the time-dependent electric field have been studied.
Tuning the Hubbard interaction strength amongst the
 states of electron-hole complexes and the parameters
 of the ac field, two strikingly different kinds of approximate qubit can be constructed within the eight trion
states. The similarity and difference between the electron dynamic
entanglement and overlapping of the wavefunction with the Bell state
have been elucidated through analyzing these two kinds of
 qubit.
 \end{abstract}
\maketitle

\section{Introduction}

The electronic transport phenomena in the quantum dots have been
enriched by the introduction of the time-dependent driving external
fields\cite{platero}, partly due to the sidebands created by the
harmonic fields, and these sidebands open additional more channels
for the tunneling
 of the interacting electrons. More intriguing effect
 is the dynamic localization produced by the harmonic field, which control
 the electron to locate one of the states permanently or in a
 relatively long duration\cite{grossmann,creffield,paspalakis,liucs,ysliu,pzhang1,pzhang2,danckwerts,murgida}.
 The proper choice of the
 external harmonic field parameters can compensate the inadequacy of
 the artificial engineered nanostructures, and provides more
 possibilities to manipulate the electronic and transport properties
 of the electrons. Since the pioneering works of Loss and
 Divincenzo\cite{loss}, the interacting electrons in the quantum dots have become one of the entities in the
 realization of solid state quantum computation. Besides taking the
 electron as the quantum information processing candidate, exciton\cite{ysliu,luis} and
  even charged exciton (trions)\cite{runge,esser,peeters,combescot} have been proposed as the candidates for the
  quantum information\cite{shabaev,pochen,goldman,economou2}.

 The two interacting
electrons in coupled quantum dots have been studied
  by many authors, where dynamic localization and delocalization have been
  found\cite{creffield,paspalakis,pzhang1}, and possibility for the generation of maximum Bell state has been discussed\cite{pzhang1}.
   Hubbard-like interacting excitons in the coupled
  quantum dots driven by harmonic field have been analyzed and
  entanglement has been realized\cite{ysliu,luis}.
In a single quantum dot, manipulation of the spin of the $\Lambda$
system consisting of the spin up/down states and trion state is
performed by Economou and coauthors\cite{sophia,bayer}, where
rotation of spin is realized through an optical way, providing an
alternative way in controlling spin for quantum information
processing.
  In our work, we
 study the dynamics of the  electron-trion $X^-$
  in the coupled quantum dots,  trion
  states in our model can be formed in either the same quantum dot or spatially-separated
  dots. The manipulation of the transition dynamics among the trion states has
  been made through tuning both the Coulombic interactions among the
  electron-hole interaction and parameters from the harmonic fields.
In the Flouquet picture, by resolving the special crossing points in
quasienergies of the system\cite{pzhang1,creffield,murgida}, the
resonant tunneling/Rabi flopping within the different trion states
can be realized. Particular interesting phenomenon lies in the fact
that the trion is capable of
 oscillating in the partial states of the all trion states for a long duration of time,
 while the tunneling to the other states becomes forbidden. In these
various
 allowed-forbidden configuration, two different kinds of quantum qubit
 have been constructed, the overlapping of the wavefunction with the
  maximum Bell state and the concurrence describing the electron entanglement
 have been compared and discussed. The so-called {\sl electron entanglement dynamic single-slit} and
 {\sl double-slit}\cite{singleslit} have been  introduced to
 describe the similarity and difference between the overlapping and
 entanglement.

Our work is organized as follows, the model together with
  the brief description of the quasienergies and associated dynamics is presented in section II;
    The quantum information properties including the generation of qubit and dynamic electron entanglement
     are given in the Section III, section IV concludes our work.
\section{The model and tunneling dynamics}

To describe the electron-trion $X^-$ in the coupled quantum dots, we
introduce two quantum dots system whose energy-level detuning for
both the electron and hole is neglected. In order for simplicity,
 The single-level approximation for both the electron and hole is adopted, that
 is to say, the harmonic driving field frequency $\omega$ is assumed to be much
lower than the energy spacing between the single-particle ground
state level and the first excited state level. If the tunneling rate
in the QD systems is much larger than the electron-hole
recombination rate and the electron-phonon interaction is
neglected\cite{mtkuo,hawrylak}, Hubbard on-site interaction among
electron and hole carriers can be employed,
 thus the Hamiltonian can be written as,
\begin{eqnarray}
\label{hamiltonian}
 {\cal H}(t) &=&\sum\limits_{i,\ell,\sigma}\epsilon
_{i,\ell,\sigma}d_{i,\ell,\sigma}^{\dagger}d_{i,\ell,\sigma}+\sum_{i}W_{i,\sigma
}\left( d_{i,L,\sigma }^{\dagger}d_{i,R,\sigma }+h.c.\right)
+\sum_{i\neq j,\sigma,\sigma}\frac{U_{ij,\sigma ,\sigma}}{2}\left(
n_{i,L,\sigma }n_{j,L,\sigma }+n_{i,R,\sigma }n_{j,R,\sigma
}\right)\nonumber \\ && +\sum_{i\neq j,\sigma
,\sigma}\frac{V_{ij,\sigma ,\sigma}}{2}\left( n_{i,L,\sigma
}n_{j,R,\sigma }+n_{i,R,\sigma }n_{j,L,\sigma }\right)~,
\end{eqnarray}
where $d_{i,\ell,\sigma }^{\dagger}$ indicates creating an electron
(hole) $i$ in the $\ell$-side ($\ell$=left (right)) dot, and
$d_{i,\ell,\sigma }$,  annihilating an electron (hole) $i$ in the
$\ell$-side ($\ell$=left (right)) dot. $W_{i,\sigma }$ is the
hopping parameter of electron (hole) between the two dots.
$U_{ij,\sigma ,\sigma}$ and $V_{ij,\sigma ,\sigma}$ denote the
intra-dot and inter-dot Coulombic interaction among the carriers
respectively. $n_{i,\ell,\sigma }$ is the number of electrons or
hole in the left (right) dot. When the QDs are biased by a harmonic
field, their energy levels will be shifted as follows,
$\epsilon^{\mp}_{i,\sigma }\left( t\right) =\epsilon _{i,\sigma
}^{0}\mp \frac{\phi}{2}\cos \left( \omega t\right)$, where the
$\phi$ is the potential proportional to the product of amplitude of
the harmonic field and spacing between the QD. The electron-trion
states can be described as follows in terms of the  many-particle
basis
$d^\dagger_{i_1,\ell_1,\sigma_1}d^\dagger_{i_2,\ell_2,\sigma_2}d^\dagger_{i_3,\ell_3,\sigma_3}\vert
0\rangle$, where $i,\ell,\sigma$ with the subscripts represent the
electron/hole, left/right, spin up/down respectively.
 There are 32 states all together, while only eight states survive after
 taking account of the above assumptions. These eight $X^-$ states can
 be represented as the ket states: $\left\vert {e}
{\bar e} h\right\rangle $, where the spin-up, spin-down electrons
and hole are arranged from the left to the right respectively in the
ket. By further introducing the number 0 and 1 to indicate whether
the carriers appear in the left or right QDs, the 8 electron-trion
states can be written
 as:  $\left\vert 0{\bar 0}0\right\rangle
,\left\vert 0{\bar 0}1\right\rangle ,\left\vert 0{\bar
1}0\right\rangle ,\left\vert 0{\bar 1}1\right\rangle ,\left\vert
1{\bar 0}0\right\rangle ,\left\vert 1{\bar 0}1\right\rangle
,\left\vert 1{\bar 1}0\right\rangle ,\left\vert 1{\bar
1}1\right\rangle$ (denoted as: $\vert \varphi_\alpha\rangle,
(\alpha=1,2,\cdots, 8))$ . Using these basis states, the
corresponding Hamiltonian can be obtained in a matrix form as,
\begin{eqnarray}
\label{hamiltonian}
{\cal H}(t)=\left(\begin{array}{cccccccc} -\frac{3}{2}{\cal F}(t)&W_h&W_{2e}&0&W_e&0&0&0 \\
W_h&2{\cal U}-\frac{1}{2}{\cal F}(t)&0&W_{2e}&0&W_e&0&0\\
W_{2e}&0&-\frac{1}{2}{\cal F}(t)&W_h&0&0&W_e&0
\\
0&W_{2e}&W_h&\frac{1}{2}{\cal F}(t)&0&0&0&W_e\\
W_e&0&0&0&-\frac{1}{2}{\cal F}(t)&W_h&W_{2e}&0\\
0&W_e&0&0&W_h&\frac{1}{2}{\cal F}(t)&0&W_{2e}
\\
0&0&W_e&0&W_{2e}&0&2{\cal U}+\frac{1}{2}{\cal F}(t)&W_h \\
0&0&0&W_e&0&W_{2e}&W_h&\frac{3}{2}{\cal F}(t)~
\end{array}\right),
\end{eqnarray}
where the gauge transformation
$\exp(-i(2\epsilon_{e\sigma}+\epsilon_{h\sigma}+U_{1h\sigma}))$ has
been performed to eliminate the
 common term in the diagonal part of the Hamiltonian; the interaction strength $|{\cal U}_1|$ and $|V_1|$ among the
  carriers in the intra-dot and inter-dot are assumed respectively,
without loss of generality. In the above effective Hamiltonian,
 ${\cal F}(t)=\phi\cos(\omega t),$ and ${\cal U}=-|{\cal U}_1|+|V_1|$. The above system can not be
  solved in an exact way due to the non-communicativity of
the Hamiltonian at different times. However Floquet theorem provides
a solution clue to the time-periodic system $:
 \left\vert \Psi _{\alpha ,m}(t)\right\rangle =\exp
\left( -i\varepsilon _{\alpha ,m}t\right) \left\vert u_{\alpha
,m}(t)\right\rangle $. The Floquet state $\left\vert u_{\alpha
,m}(t)\right\rangle$ satisfies the following alternative
Schr\"odinger equation,
\begin{eqnarray}
\left( {\cal H}\left( t\right) -i\frac{\partial }{\partial t}\right)
\left\vert u_{\alpha ,m}\right\rangle =\varepsilon _{\alpha
,m}\left\vert u_{\alpha ,m}\right\rangle,
\end{eqnarray}
where the quasienergy $\epsilon_{\alpha,m}$ is restricted in the
first Brillioun zone
 $[-\frac{\omega}{2},\frac{\omega}{2}]$.
The merit of the introduction of the time-dependent external fields
lies in that
  the dynamics can be tuned to the regime of the interest. While
  from the generalized parity symmetry point of view, the subsequent Floquet states can
   be classified into even parity or odd parity
states, the quasi-energies belonging to the different and same
symmetry may develop into exact crossings and avoided crossings
respectively\cite{platero,hanggibook}. By choosing the parameters
corresponding to these special points, the intriguing quantum
information properties will be disclosed, which have been reported
in the section III.

The changing behavior of the quasienergies and minimum propagator
probability $P^{min}_\alpha(\equiv\min(|c_\alpha(t_j)|^2),
j=1,2,\cdots, N,$ with $t_N=30T (\alpha=1,2))$ with the driving
potential
 $\phi$ has been illustrated in Fig.1, where we use the parameters: $W_{e}=W_{2e}=1.0, W_{h}=0.6$, $\omega =2.0$,
 the Coulombic interaction difference ${\cal U}$ has been taken
 $-1$.
  The enhancement of the Coulombic interaction
 ${\cal U}$ induces asymmetry of quasienergies about the horizontal
 axis. Closer inspection of the Hamiltonian in Eq.(\ref{hamiltonian}) reveals
 that the present system can be viewed as the two coupled dot arrays with four
 dots(states) in each array, and hopping term between the array is $W_e$ (see the Fig.2).
 This explains why the quasienegies are split into two parts. Also in Fig.1, the survival
 probabilities $P^{min}_1$ and $P^{min}_2$ have been depicted, where
 the initial conditions have been chosen as $c_1(0)=1.0$ or $c_2(0)=1.0$
  respectively. It is interesting that the peaks for $P^{min}_1$ and
  $P^{min}_2$ appear in different exact crossing points of the
  quasienergies. All these show that the
  states: $\vert \varphi_1\rangle$ and $\vert \varphi_2\rangle$
  contain different generalized symmetry component of Floquet functions\cite{holthaus2,hanggibook}.
  Another interesting phenomenon is that
  $P^{min}_2$ have two neighboring peaks, forming
  bistable structures which result from the closely-spacing exact crossings in
  the quasienergies.

\section{The generation of qubit and dynamic electron entanglement}

 Rajagopal and Rendell analyzed eight states from triple-qubit,
 where robust and fragile entanglement are classified\cite{rajagopal}.
 Although there are eights states in our system, these eight states from three
non-identical particles can not be mapped onto triple-qubit eight
states, since there exists
 spatial freedom as well as spin freedom in our system.
 The states $\vert 0{\bar 0}0\rangle$ and $\vert 1{\bar 1}1\rangle$
 represent that the trion stays in the same QD, while the other
 states belong to the spatially separated trion states.

 Even if
 there is no obvious qubit  in our system, we can
 tune the system parameters and harmonic driving field to
 create the qubit.
 Fig.3 gives transition dynamics among the
 eight trion states, where parameters used are: ${\cal U}=-20.0, W_h=0.6,
 W_e=1.7, \omega=2.0, \phi=24.6$. The driving potential $\phi$ is chosen to
 be around
 the avoided crossing point of the quasienergies. From the transition dynamics pattern in the figure, it
 can be clearly seen that the quantum transition nearly takes place among
 only
  three states: $|0{\bar 0}0\rangle, |0{\bar 1}0\rangle,$ and $|1{\bar
  0}0\rangle$. Particularly interesting phenomenon lies in that the
  dynamics for $|0{\bar 1}0\rangle,$ and $|1{\bar
  0}0\rangle$ is highly equivalent in both the amplitude and
  phase. In this circumstance, $|0{\bar 0}0\rangle$ together with the normalized linear combination
  state:
   $\vert \eta\rangle\equiv\frac{1}{\sqrt{2}}(|0{\bar 1 }0\rangle+|1{\bar 0}0\rangle)$, can be esteemed as a
    spatial-separated qubit\cite{doll}:
    $\vert \gamma\rangle \approx(c_1(t)|0{\bar 0}0\rangle+c_3(t)|\eta\rangle)$.

 Inspection of the configuration of the eight trion states
 indicates that electron dynamic entanglement can be realized among the eight trion
states from spatial degree of freedom.
 In order to measure the degree of the entanglement, we need to calculate
 the concurrence\cite{rajagopal,wooters1} of the electron-electron density
 matrix, which is defined as,
\begin{eqnarray}
\varrho_{ee}:={\textrm Tr}_h \varrho_{eeh}=
\sum\limits_1^{N_h}\left({\bf I}_{ee}\bigotimes{_h}\langle
\chi_j|\right) \varrho_{eeh}\left({\bf
I}_{ee}\bigotimes|\chi_j\rangle_h\right)~,
\end{eqnarray}
where ${\textrm Tr}_h$ represents the partial trace over the hole
basis $|\chi_j\rangle_h$; $\bigotimes$ is the direct product and
${\bf I}_{ee}$ is the $4\times 4$ unit matrix.
 From the above equation, the electron-electron density matrix $\varrho^{ee}$ can be evaluated to be
\begin{eqnarray}
\label{rhoeq}
\varrho_{ee}=\left(\begin{array}{cccc} |c_1|^2+|c_2|^2 &c_1c^*_3+c_2c^*_4 &c_1c^*_5+c_2c^*_6&c_1c^*_7+c_2c^*_8 \\
c_3c^*_1+c_4c^*_2& |c_3|^2+|c_4|^2&c_3c^*_5+c_4c^*_6&c_3c^*_7+c_4c^*_8\\
c_5c^*_1+c_6c^*_2&c_5c^*_3+c_6c^*_4&|c_5|^2+|c_6|^2&c_5c^*_7+c_6c^*_8\\
c_7c^*_1+c_8c^*_2&c_7c^*_3+c_8c^*_4&c_7c^*_5+c_8c^*_6&|c_7|^2+|c_8|^2\\
\end{array}\right)
\end{eqnarray}
The other matrix ${\widetilde{\varrho}}_{ee}$, necessary for the
calculation of the concurrence can be constructed through the
following way,
\begin{eqnarray}
\label{rhoeqt}
~&&{\widetilde{\varrho}}_{ee}=\left(\sigma_y\bigotimes\sigma_y\right)
{\varrho}_{ee}^*\left(\sigma_y\bigotimes\sigma_y\right)\nonumber
\\
&&=\left(
\begin{array}{cccc}
|c_7|^2+|c_8|^2 & -(c_7c^*_5+c_8c^*_6) & -(c_7c^*_3+c_8c^*_4)&
c_7c^*_1+c_8c^*_2\\
-(c_5c^*_7+c_6c^*_8)& |c_5|^2+|c_6|^2& c_5c^*_3+c_6c^*_4 &
-(c^*_5c_1+c_6c^*_2)\\
-(c_3c^*_7+c_4c^*_8)& c_3c^*_5+c_4c^*_6 & |c_3|^2+|c_4|^2 &
-(c^*_3c_1+c_4c^*_2)\\
c_1c^*_7+c_2c^*_8 & -(c_1c^*_5+c_2c^*_6) &-(c_1c^*_3+c_2c^*_4)&
|c_1|^2+|c_2|^2\\
\end{array}
\right)~.
\end{eqnarray}
The procedure for the calculation of the concurrence is to find the
eigenvalues of the Hermitian matrix
$\sqrt{\sqrt{\varrho}\widetilde{\varrho}\sqrt{\varrho}}$. The
concurrence is obtained as
$C_{ee}=\max\{\lambda_1-\lambda_2-\lambda_3-\lambda_4,0\}$, where
$\lambda_i~(i=1,2,3,4)$ are the square roots of the eigenvalues of
the matrix $\sqrt{\sqrt{\varrho}\widetilde{\varrho}\sqrt{\varrho}}$
arranged in decreasing order\cite{wooters1,rajagopal}.

  In the case of Fig.3, let us focus on the electron entanglement of the state $|0\bar{1}0\rangle$ and
 $|1\bar{0}0\rangle$, the non-vanishing $c_i(t)$ are $c_1(t),~c_3(t),~c_5(t)$
  with the other $c_i(t)~(i=2,4,6,7,8)$ almost vanishing, leading to
\begin{eqnarray}
\label{occurence}
~\varrho\widetilde{\varrho}\approx\left(\begin{array}{cccc}
 0 & 2c_1^*c_3|c_5|^2 & 2c_1^*c_5|c_3|^2 &-2c^*_5c_3|c_1|^2\\
 0 & 2|c_3|^2|c_5|^2 &2c_3^*c_5|c_3|^2& -2c^*_1c_5|c_3|^2 \\
 0 & 2c^*_5c_3|c_5|^2 & 2|c_3|^2|c_5|^2& -2c^*_3c_1|c_5|^2\\
 0&0&0&0\\
\end{array}\right)~,
\end{eqnarray}
whose eigenvalues can be evaluated to be
$\{4|c_3|^2|c_5|^2,0,0,0\}$. Hence, the concurrence
$C_{ee}(t)=2|c_3(t)||c_5(t)|$, which was depicted in the up panel of
Fig.4. The overlapping of the state $|\psi(t)\rangle$ with the
 maximum-entangled Bell state
 $|\psi_{Bell}\rangle=\frac{1}{\sqrt{2}}(|0{\bar 1}0\rangle + |1{\bar
 0}0\rangle)$, i.e., $\rho_1(t)=\vert\langle \psi(t)|\psi_{Bell}\rangle\vert^2\approx ({|c_3(t)+c_5(t)|^2})/{2}$ is
 given in the middle panel of the Fig.4, and the bottom panel gives
 both the curves for comparison.  From the figures, it can be clearly seen that
 the occurrence oscillates with time and reach the maximum
 entanglement when the quenching of the trion state $|0{\bar 0}0\rangle$ in the left
 QD occurs. Particularly, it is remarkable that the concurrence is coincident with
  the overlapping in the present case. This coincidence can be
  attributed to the fact that the propagator $c_3(t)$ is equal to
  $c_5(t)$, leading to $\rho_1(t)\approx|c_3(t)+c_5(t)|^2=2|c_3(t)||c_5(t)|\approx
  C_{ee}(t)$. The situation here can be pictorially described as the
  {\sl electron entanglement dynamic single-slit}, where the maximum Bell state passes through
  the qubit  $\vert \gamma\rangle \approx(c_1(t)|0{\bar 0}0\rangle+c_3(t)|\eta\rangle)$. The qubit here only has a
  dynamic single-slit $c_3(t)|\eta\rangle$ (the other component of qubit $c_1(t)|0\bar{0}0\rangle$ is blind to the Bell state), which is the very reason
  why overlapping is equivalent to the concurrence.

   The full dynamic
  entanglement of the electrons taking account of all propagators $c_i(t),~(i=1,2,\cdots,8)$ is
  depicted in Fig. 5, where both the
  concurrence $C_{ee}(t)$ and overlapping $\rho_1(t)$ is given for
  comparison. The difference between the full concurrence and
  overlapping
   mainly results from the leaving out the tiny but non-vanishing propagator
  $c_7(t)$. The presence of the tiny propagator $c_7(t)$ induces the
   other entanglement: $\frac{1}{\sqrt{2}}(|0{\bar 0}0\rangle$+$|1{\bar 1}0\rangle)$.
   The full dynamic concurrence is composed of two sorts of
   entanglement $\frac{1}{\sqrt{2}}(|0{\bar 1}0\rangle+|1{\bar 0}0\rangle)$,
   and $\frac{1}{\sqrt{2}}(|0{\bar 0}0\rangle$+$|1{\bar
   1}0\rangle)$, but the former one is much larger than the latter.
   This explains why the full concurrence is slightly larger than
   the overlapping (see the bottom panel of Fig.5).

Does the {\sl dynamic double-slit} for the electron
  entanglement exist? As matter of fact, besides the above mentioned qubit, there exists
 other form of qubit which can be considered as electron entanglement dynamic double-slit,
  if we alter both the Coulombic
 interaction strength and harmonic driving field parameters in a proper way.
 The case is depicted in Fig.6, where the strong driving potential
 $\phi=40.7$ are used, the other parameters used are the same as those in
  Fig.1. The transition dynamics shown in the figure is calculated
  with the initial condition: $(c_1(0),c_2(0),c_3(0),c_4(0),c_5(0),c_6(0),c_7(0),c_8(0))=(1/\sqrt{2},0,0,0,0,0,1/\sqrt{2},0)$.
 It is evident that the transition dynamics for $|0{\bar 0}0\rangle, |1{\bar1}0\rangle$
  is virtually the same, making transition to $|0{\bar 0}1\rangle,  |1{\bar 1}1\rangle$
  respectively in a nearly synchronous way. The outcome can
  be attributed to the fact that the strong driving force prevails
  over the attractive force between the electrons and hole,
  making the hole in one QD tunnel into the other QD.
  These four trion states are able to forming new approximate qubit: $\vert \delta\rangle\approx
  (c_1(t)\vert \alpha\rangle + c_2(t)\vert \beta\rangle)$, where $|\alpha\rangle=\frac{1}{\sqrt{2}}(|0{\bar
0}0\rangle+|1{\bar1}0\rangle)$
  and $|\beta\rangle=\frac{1}{\sqrt{2}}(|0{\bar 0}1\rangle+|1{\bar 1}1\rangle)$.
   The difference between the qubit $\vert\delta\rangle$ and the previously defined
   qubit $\vert \gamma\rangle$ lies in the different degree of spatial separation of the electrons within the
    two components of qubit. In the previously defined qubit $\vert\gamma\rangle$, only
    one electron populates in the different QD within the two
    components of qubit; while all the two electrons populate in different QD
    in the present qubit $\vert\delta\rangle$.

 Through taking inner product with the Bell state $\vert
 \beta\rangle$, the probability $\rho_2(t)=|\langle \beta|\Psi(t)\rangle|^2$ for the system to stay in the Bell
 state $\vert
 \beta\rangle$ is given in the middle panel of Fig.7, where the up panel gives the concurrence
  $C_{ee}(t)\approx 2|c_2(t)||c_8(t)|$, when we force the propagator $|c_1(t)|^2$ and $|c_7(t)|^2$ to vanish . The bottom panel gives both concurrence and overlapping
  for comparison, where the envelope of the overlapping is almost
  coincident with the concurrence.
 But there is striking difference
 between the concurrence and overlapping, the pattern in the
 latter figure is irregular, forming many spikes. The reason behind
 this irregularity comes from the tiny phase difference between the
 two propagators $\vert c_2(t)\vert^2$, and $\vert c_8(t) \vert^2$, which
 are illustrated in the Fig.8.
 The quantum interference from these two probability terms leads to the
 shaping of the irregular pattern in Fig.7.  Similar to Fig.7, Fig.9
 gives the overlapping $\rho_3(t)=|\langle \alpha|\Psi(t)\rangle|^2$ and concurrence $C_{ee}(t)\approx
 2|c_1(t)||c_7(t)|$ where both $c_2(t)$ and $c_8(t)$ have been set to
  vanish. The envelope of the overlapping matches the concurrence in
  a perfect way. As matter of fact, both Fig.7 and Fig.9 correspond
  to the case of shutting down one of the two dynamic slits
  $c_1(t)|\alpha\rangle$ and $c_2(t)|\beta\rangle$.
It should be pointed out that the state of the hole play the role of
picking out one of the two the Bell states $|\alpha\rangle$ or
$|\beta\rangle$. This selecting process corresponds to the
overlapping of the wavefunction with state $|\alpha\rangle$ or
$|\beta\rangle$.

   When both the dynamic slit become transparent, the full
   concurrence can be approximately calculated as
\begin{subeqnarray}
&&C_{ee}(t)\approx\sqrt{A+\sqrt{BC}}\\
&&A=\left[\left(|c_1(t)|^2+|c_2(t)|^2\right)\left(|c_7(t)|^2+|c_8(t)|^2\right)
+|c^*_1(t)c_7(t)+c^*_2(t)c_8(t)|^2\right]\\
&& B=\left(|c_7(t)+c_8(t)|^2\right) {\textrm
Re}\left(c_1^*(t)c_7(t)+c^*_2(t)c_8(t)\right)\\
&& C=\left(|c_1(t)+c_2(t)|^2\right) {\textrm
Re}\left(c_1^*(t)c_7(t)+c^*_2(t)c_8(t)\right)~,
\end{subeqnarray}
where ${\textrm Re}$ represents the real part.
   The exact numerical value of the full concurrence $C_{ee}(t)$ is plotted
   in the up panel of Fig.10, the bottom
   panel of the same figure gives  $\rho_2(t)$, $\rho_3(t)$ and $C_{ee}(t)$ for
   comparison, where the full concurrence is
   coincident with the envelope of either $\rho_2(t)$ or $\rho_3(t)$ in different moment.
    The concurrence from dynamic double-slit does own different evolution behavior than the wavefunction
    does in quantum mechanics.
In contrast to the usual double-slit, where the wavefunctions
experience the prominent quantum interference, the terminology
dynamic single-slit and double-slit here are for pictorial
 description since the entanglement interference here bears the striking
 different behavior than the usual double-slit wavefunction
 interference as illustrated in Fig.5 and Fig.10.

\section{Concluding remark}
We have investigated the coherent transition dynamics and explored
the quantum information properties of the electron-trion $X^-$
confined in two coupled quantum dots driven by an ac electric field.
 The transition dynamics among
the eight trion states is controlled through the consideration of
the symmetry of the Floquet states, which is embodied in the avoided
crossing and exact crossing in the quasienergies spectrum. The
tunneling dynamics among the eight trion states also depends
 sensitively on the difference of Coulombic interactions between the
 electrons and hole in the same QD and spatially-separated QD.
Tuning the motion of the electrons and hole to evolve in partial
states of the whole eight trion states, the allowed and forbidden
trion states within the transition dynamics can be captured through
which the two different kinds of the qubit can be constructed. From
spatial degree of freedom, electron entanglement can be produced and
 the resulting concurrence has been given in both analytical and exact numerical
results. The similarity and difference between
 the concurrence and overlapping of the wavefunction with the maximum Bell state have been
  elucidated. Pictorially, one kind of qubit can be conceived as dynamic single-slit, while
  the other kind of qubit can be considered as the the
dynamic double-slit for the electron entanglement.

\begin{acknowledgments}
This work is  supported by the Shanxi Liuxue Foundation, and  partly
supported by Natural Science Foundation of China with No. 10475053.
\end{acknowledgments}

\newpage
\centerline{\includegraphics[width=7.00in,height=3.8in]{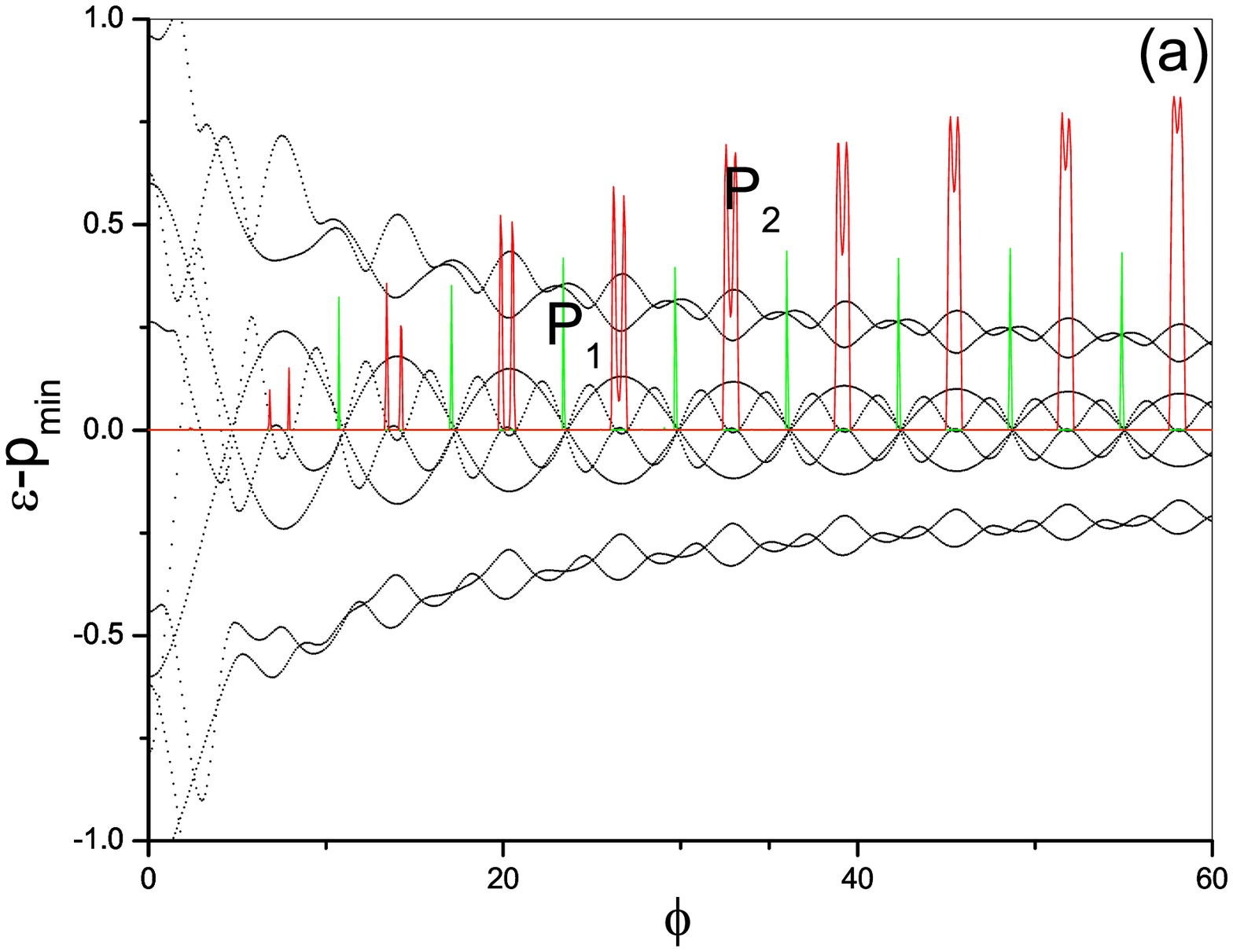}}

\makeatletter\def\@captype{figure}\makeatother\caption{Quasienergies
$\epsilon$ and $P^{min}_\alpha(\phi), (\alpha=1,2)$ plotted as the
external ac driving potential $\phi$, the minimum propagator
probabilities $P^{min}_\alpha(\phi)$ in a 30 periods time duration
experience completely different tunneling behavior for $\alpha=1$,
and $2$. }

\centerline{\includegraphics[width=3.00in,height=2.0in]{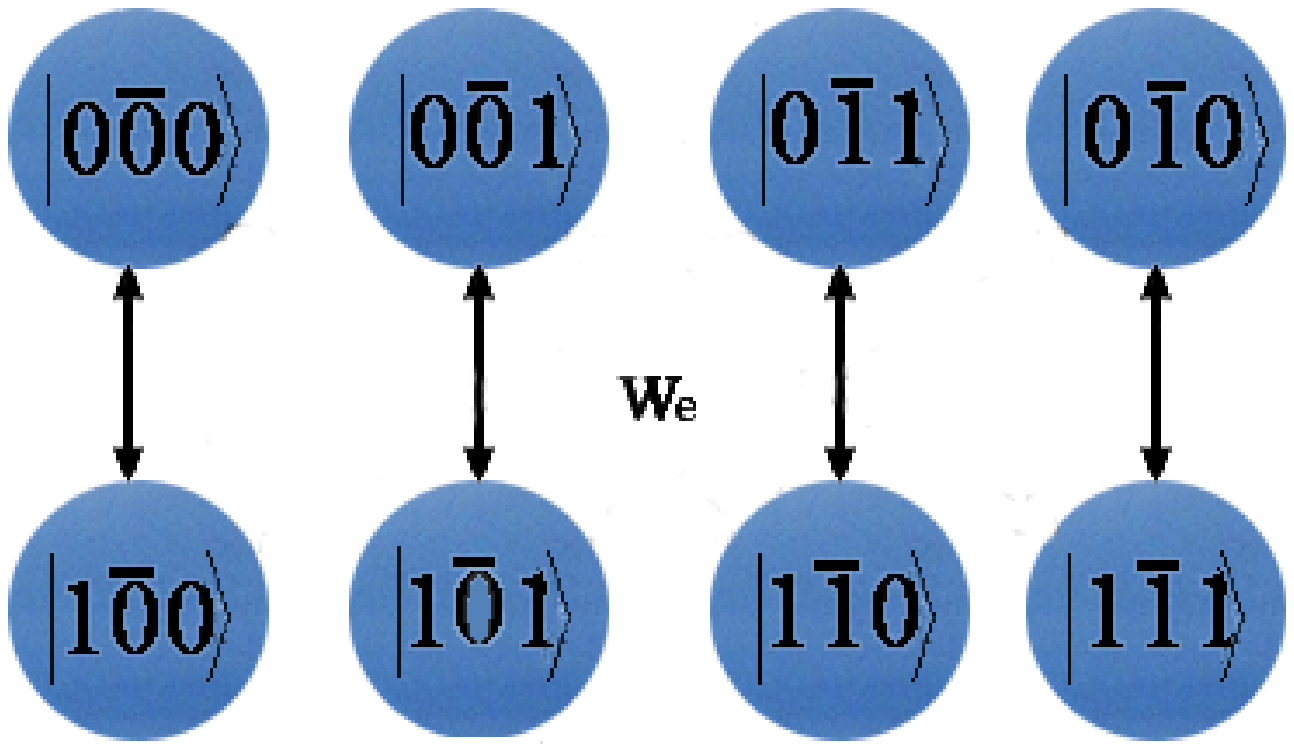}}
\makeatletter\def\@captype{figure}\makeatother\caption{The
equivalent schematic
 picture of the Hamiltonian defined in Eq.(\ref{hamiltonian}), resembling the two arrays of single-level dots
 (states) coupled through the hopping term $W_e$.
  }
\centerline{\includegraphics[width=7.00in,height=4.0in]{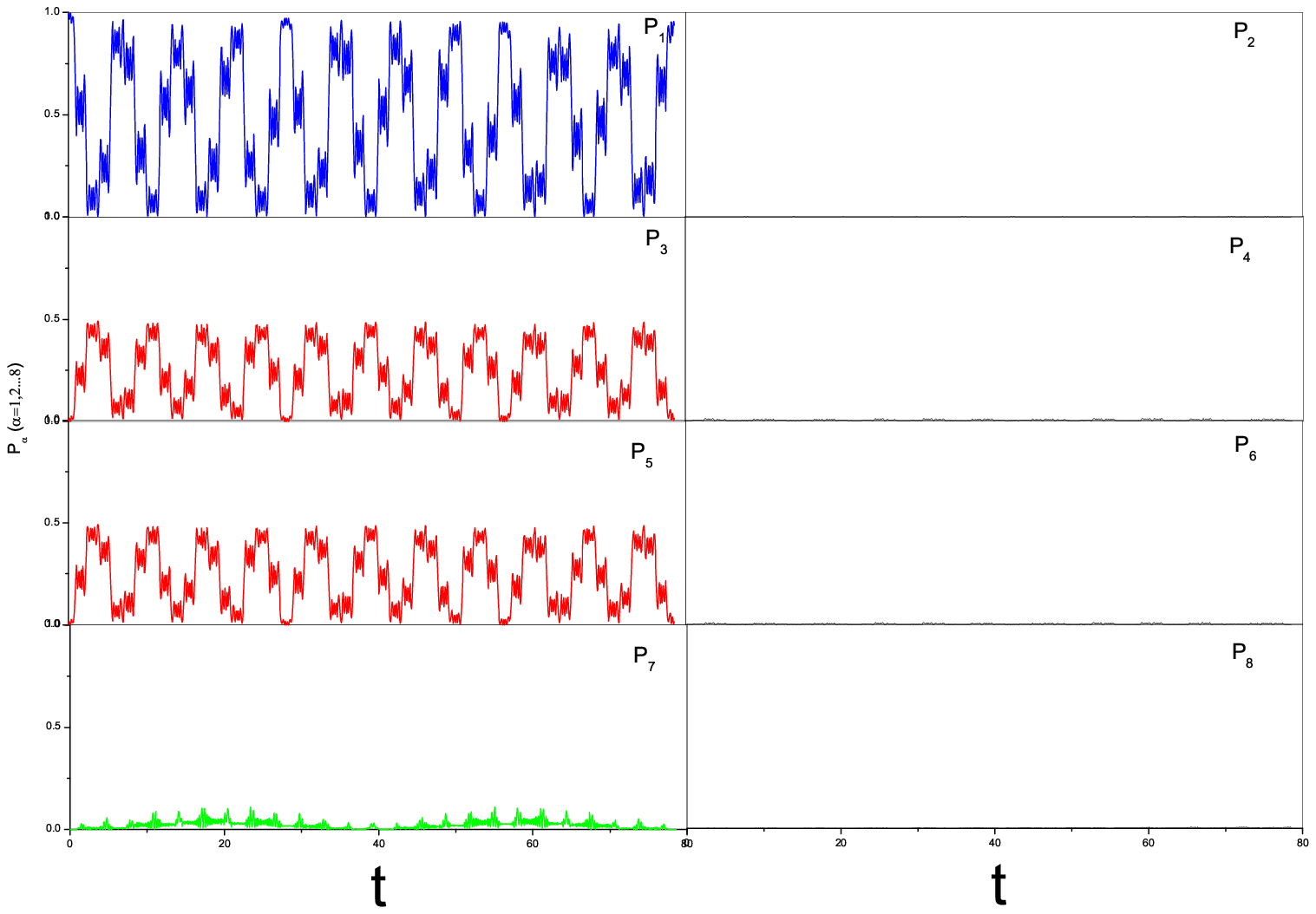}}
\vspace{-1.5cm}
\makeatletter\def\@captype{figure}\makeatother\caption{The quantum
transition dynamics among the eight trion states leading to the
formation of the first kind of approximate dynamic qubit, the
parameters used are declared in the text.}

\centerline{\includegraphics[width=7.00in,height=3.8in]{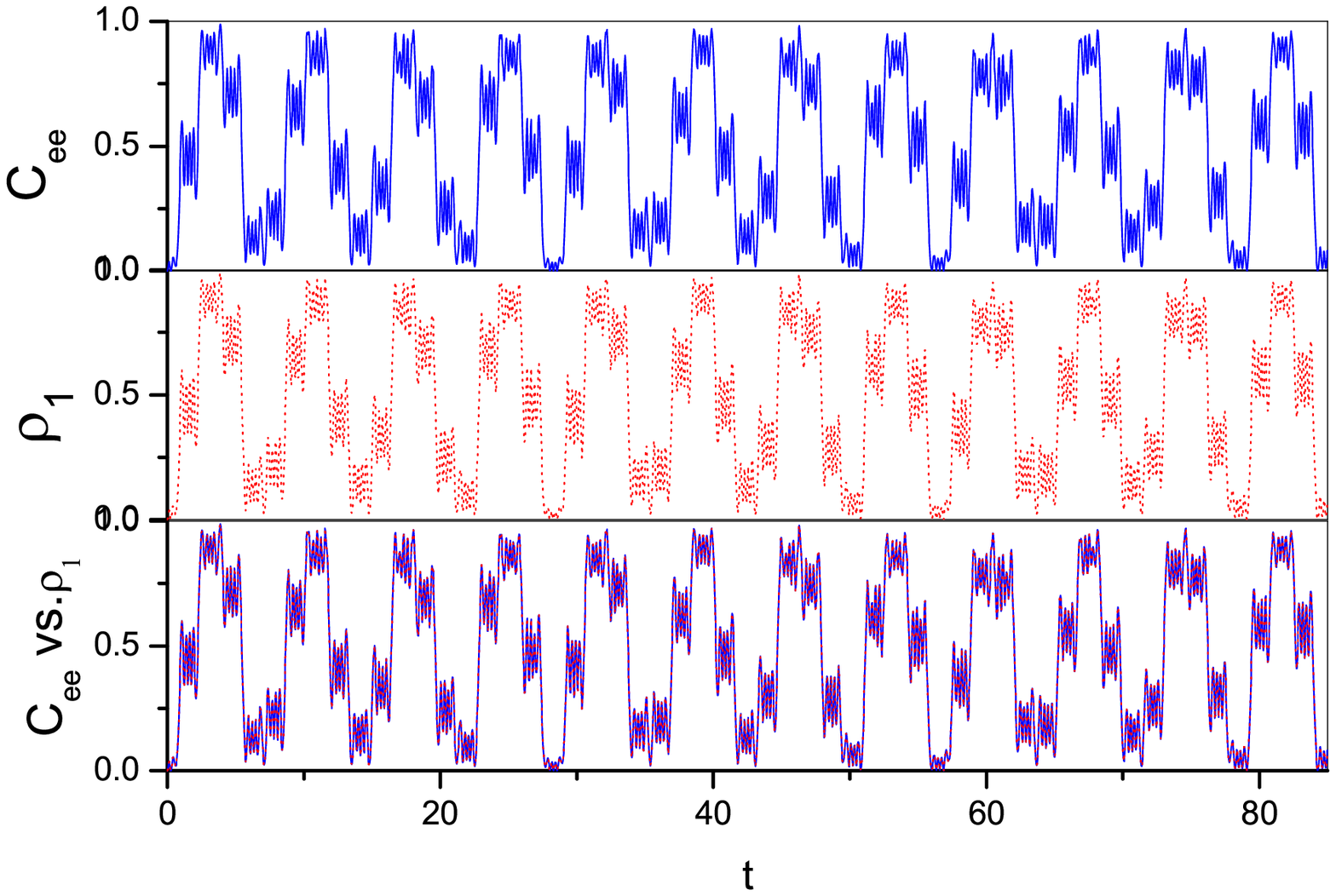}}
\vspace{-1.5cm}
\makeatletter\def\@captype{figure}\makeatother\caption{The
approximate concurrence calculated from the analytical expression
$C_{ee}(t)=2|c_3(t)||c_5(t)|$ and exact numerical overlapping of the
wavefunction with the maximum Bell state, showing the consistency of
the two quantity, the parameters used are the same as those in
Fig.3.}

\centerline{\includegraphics[width=7.00in,height=3.8in]{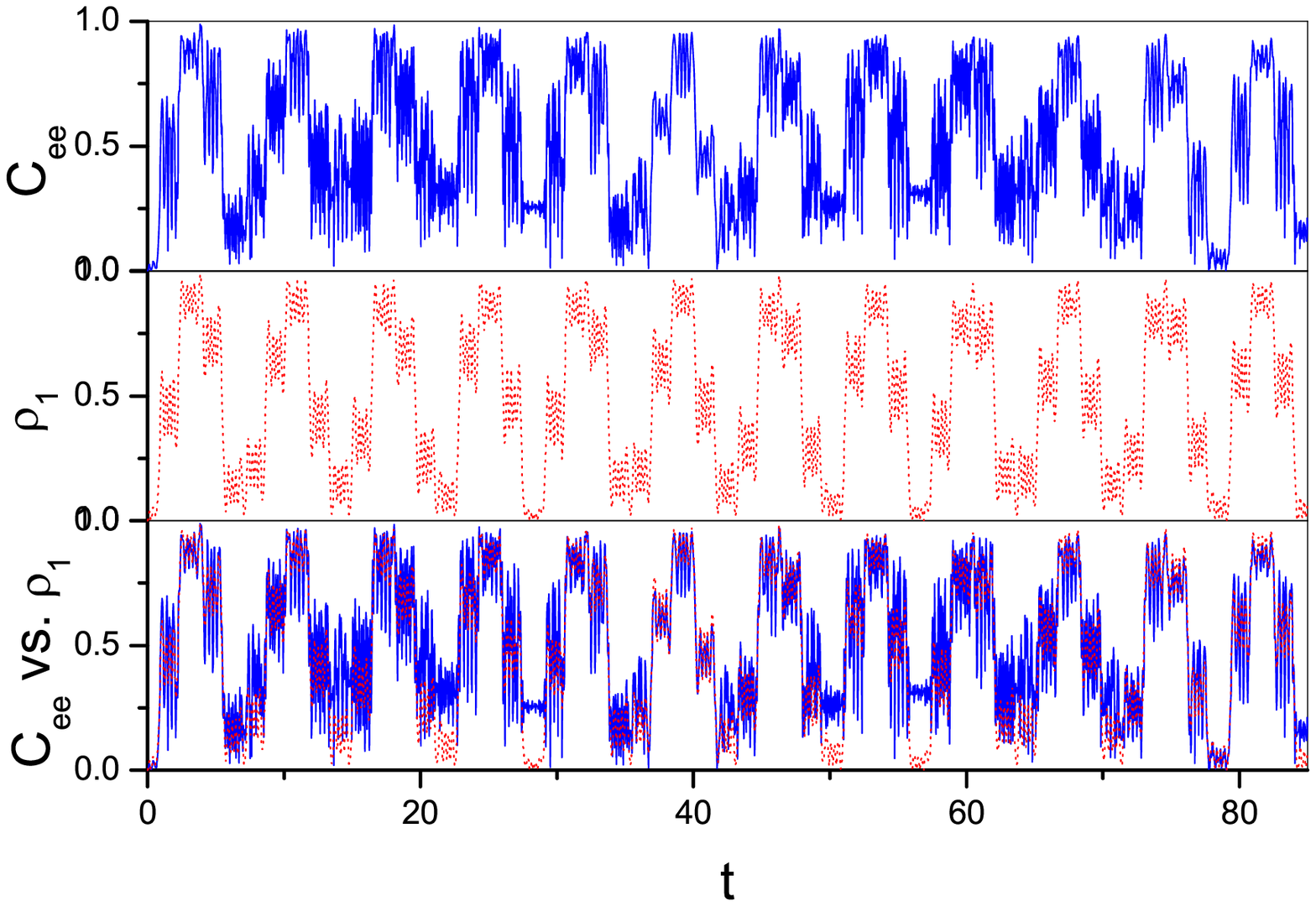}}
\vspace{-1.5cm}
\makeatletter\def\@captype{figure}\makeatother\caption{The same as
Fig.4, except that the concurrence is calculated from the exact
numerical result taking account of all $c_i(t),~ (i=1,2,\cdots,8)$.
}

\centerline{\includegraphics[width=7.00in,height=3.8in]{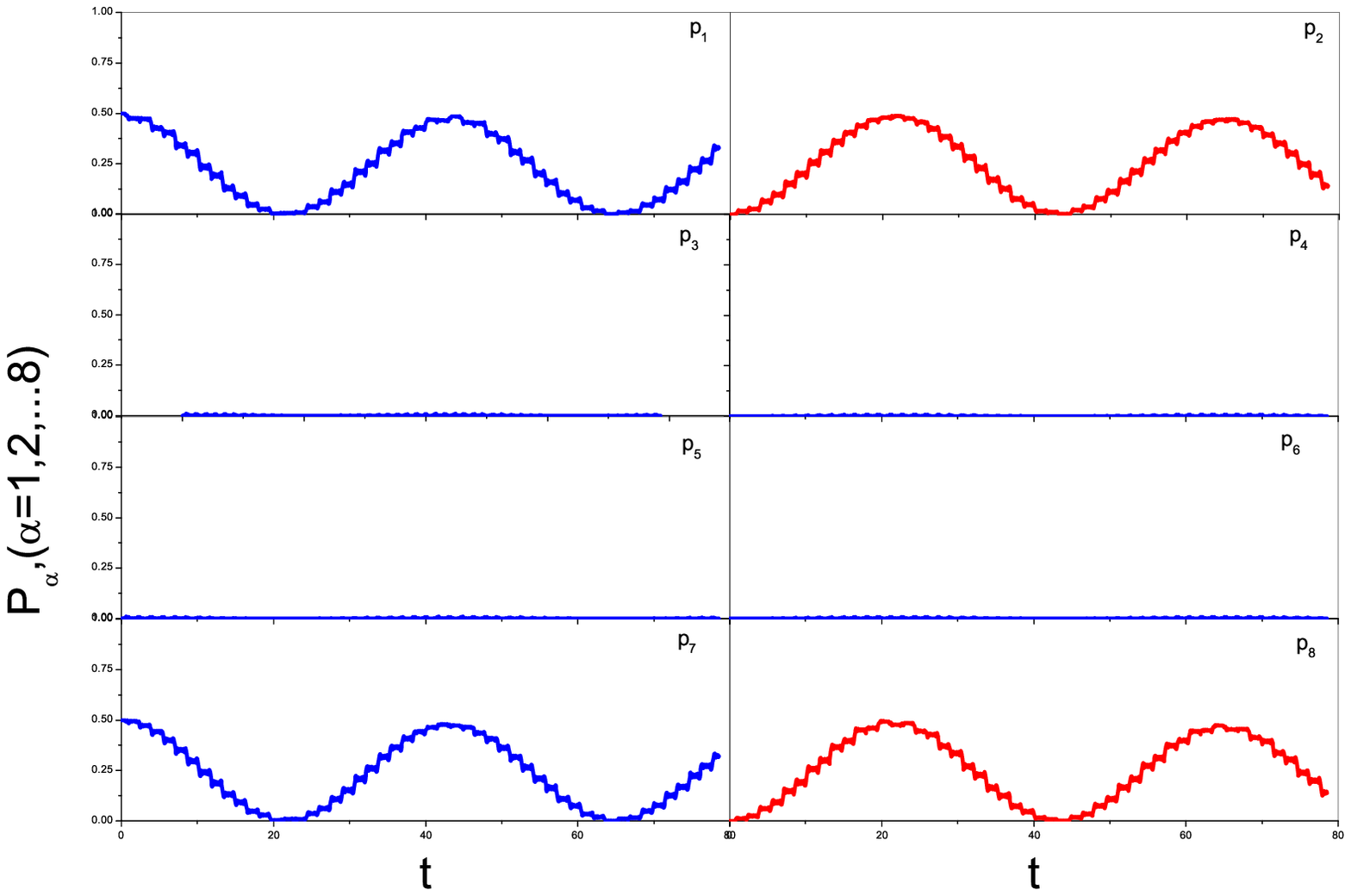}}
\vspace{-1.5cm}
\makeatletter\def\@captype{figure}\makeatother\caption{The quantum
transition dynamics among the eight trion states leading to the
formation of the second kind of approximate dynamic qubit, the
parameters used are declared in the text. }

 \centerline{\includegraphics[width=7.00in,height=3.8in]{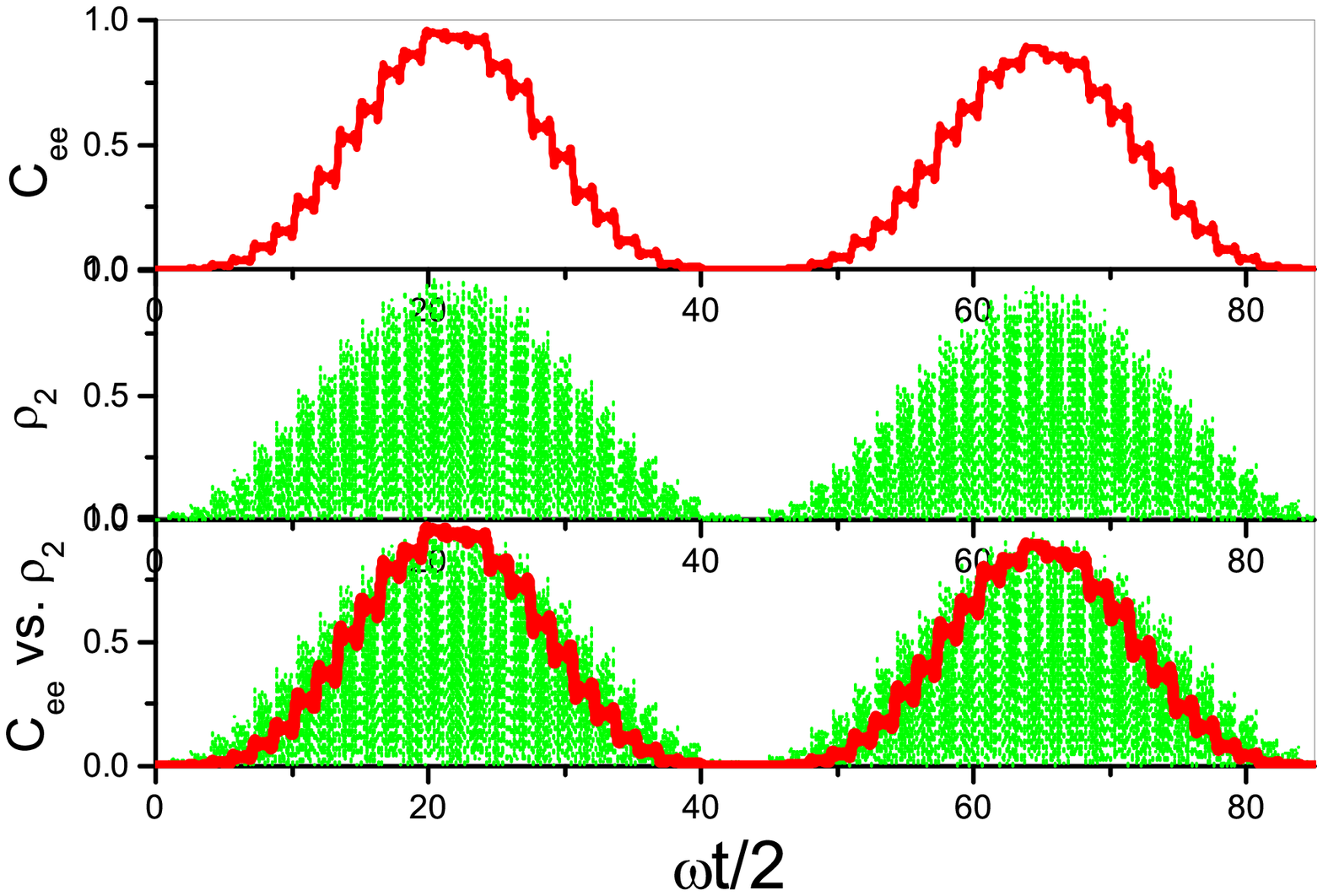}}
 \vspace{-1.5cm}
 \makeatletter\def\@captype{figure}\makeatother\caption{The approximate concurrence calculated from $C_{ee}$
  with $c_1(t)$ and $c_7(t)$ being set to vanish (we weighted $c_i(t), (i=1,3,5,7)$ by multiplying
  $\sqrt{|c_1(t)|^2+|c_3(t)|^2+|c_5(t)|^2+|c_7(t)|^2}$, and $c_i(t), (i=2,4,6,8)$ by multiplying
  $\sqrt{|c_2(t)|^2+|c_4(t)|^2+|c_6(t)|^2+|c_8(t)|^2}$), versus the exact numerical overlapping of the wavefunction the maximum Bell state,
   the envelope of the overlapping is coincident with the concurrence.}
\centerline{\includegraphics[width=7.00in,height=3.8in]{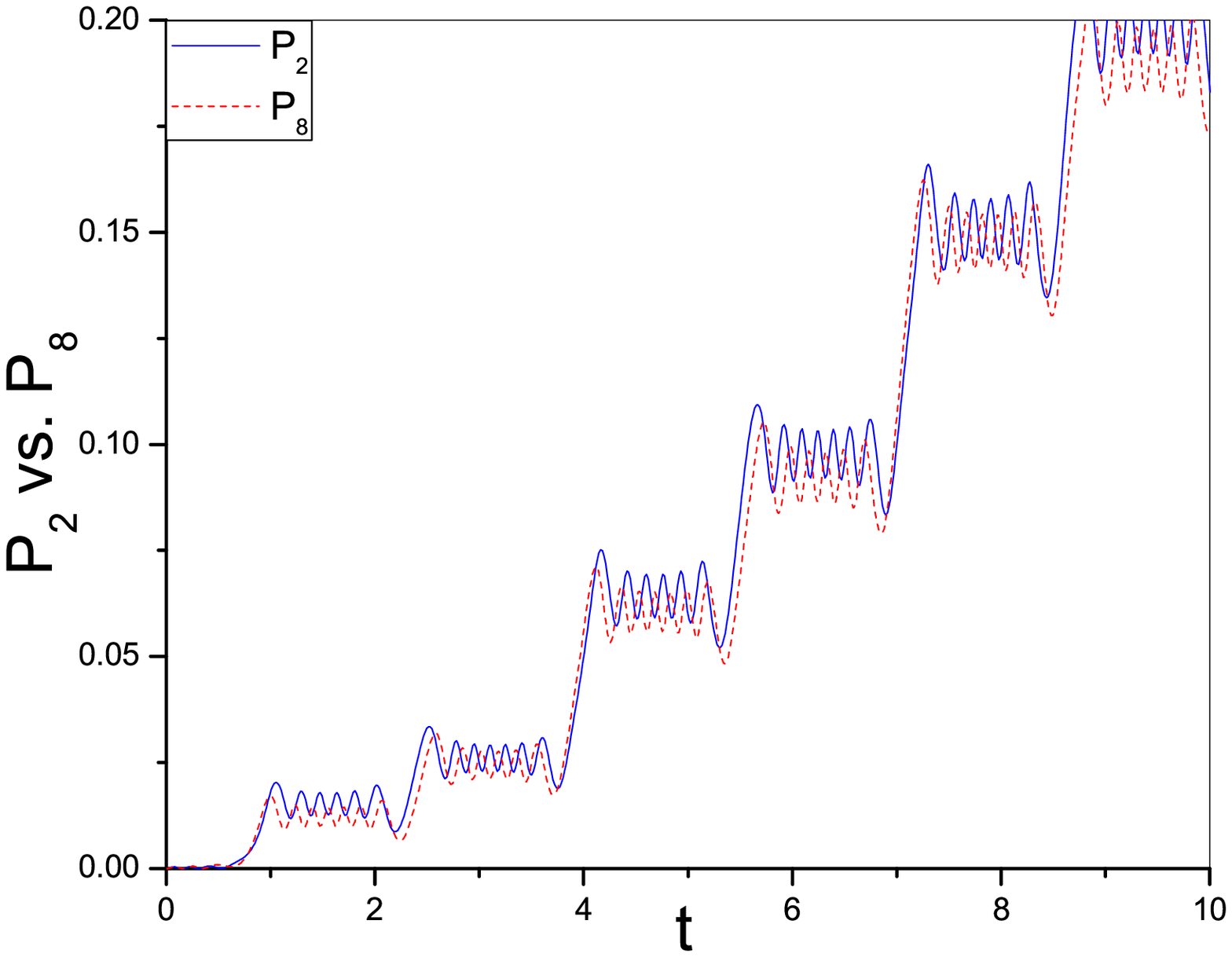}}
\vspace{-1.0cm}
 \makeatletter\def\@captype{figure}\makeatother\caption{The propagator $P_2=|c_2(t)|^2$ plotted with blue solid line and  $P_8=|c_8(t)|^2$ plotted with
  red dashed line, showing the tiny phase difference between the two propagators probability, leading to the spikes of the overlapping in Fig.7.}
\centerline{\includegraphics[width=7.00in,height=3.8in]{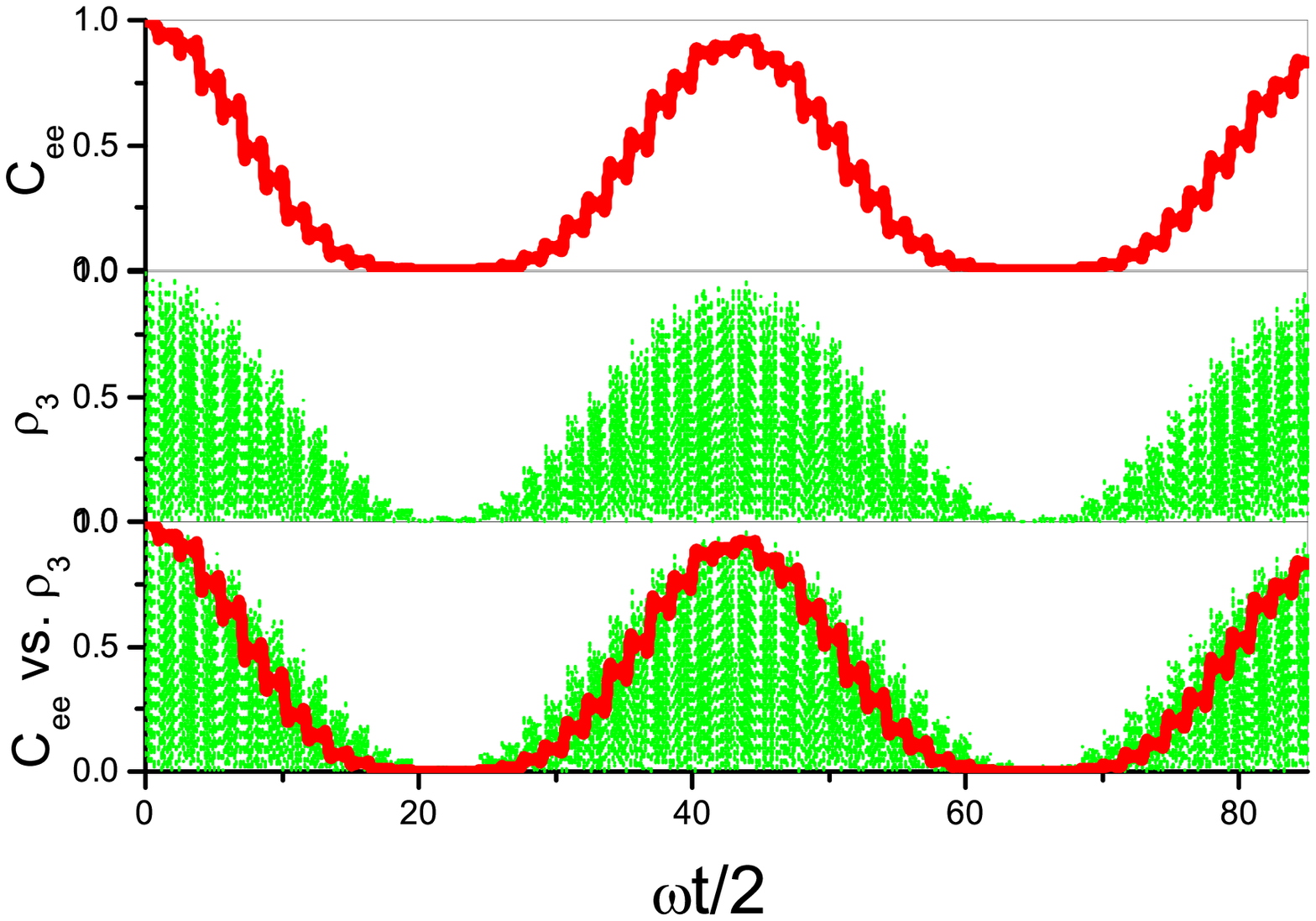}}
\vspace{-1.5cm}
 \makeatletter\def\@captype{figure}\makeatother\caption{The approximate concurrence calculated from $C_{ee}$
  with $c_2(t)$ and $c_8(t)$ being set to vanish (we use the same weight as Fig.7), and exact numerical overlapping of the wavefunction the maximum Bell state,
   the envelope of the overlapping is coincident with the concurrence. }
\centerline{\includegraphics[width=7.00in,height=3.8in]{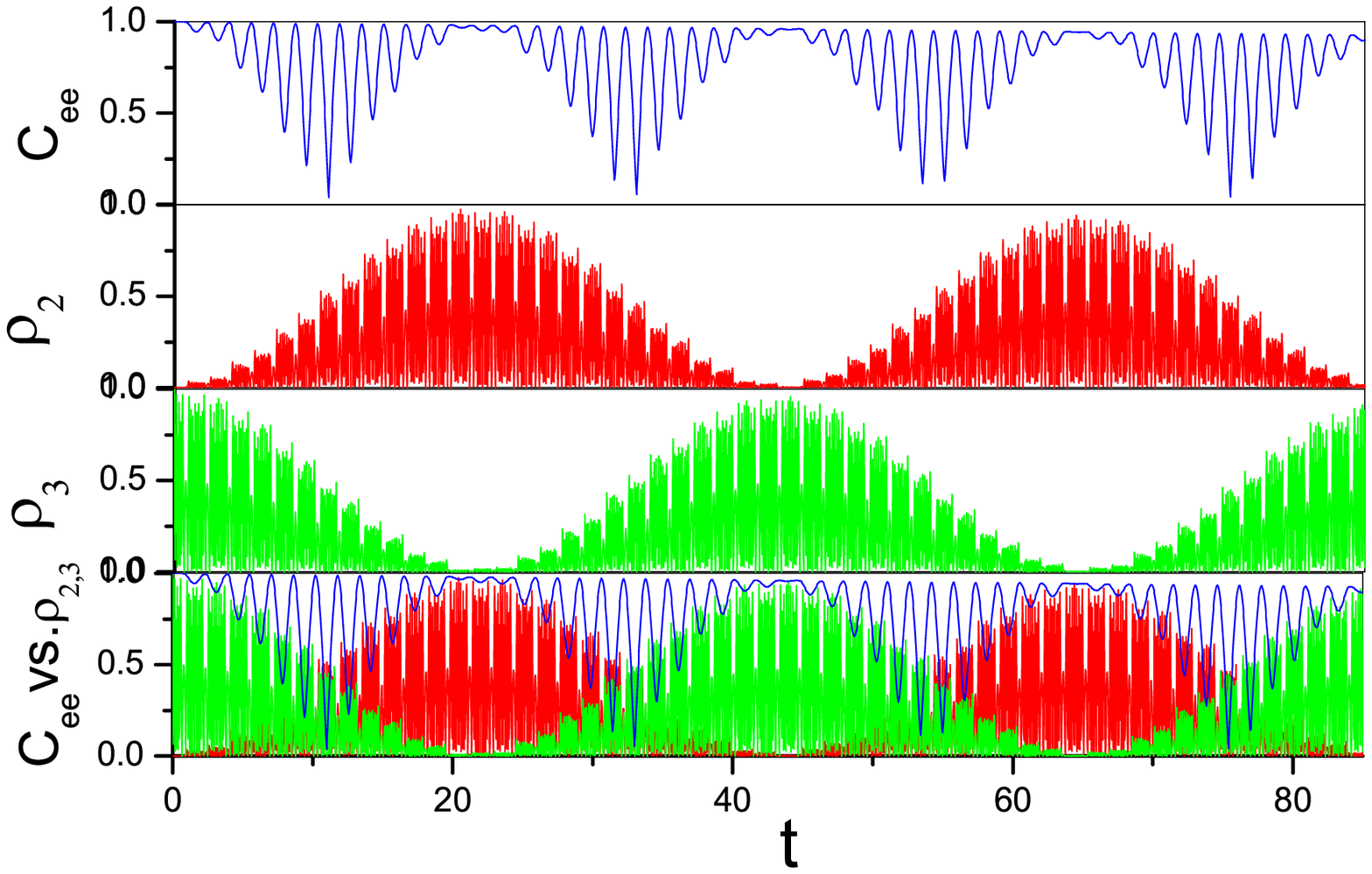}}
\vspace{-1.5cm}
 \makeatletter\def\@captype{figure}\makeatother\caption{The full concurrence calculated from the exact
numerical result taking account of all $c_i(t),~ (i=1,2,\cdots,8) $
is plotted against the two overlapping, showing the striking
difference between the full concurrence and individual overlapping
in dynamic double-slit for the electron entanglement. }


\begin{thebibliography}{99}
\bibitem{singleslit} The electron entanglement dynamic single slit and double slit
are for pictorial and physical description, the former means that
there is only one mechanisms contribution to the concurrence, while
the latter contains two mechanisms.
\end{thebibliography}
\end{document}